\documentclass[prl,aps,twocolumn,longbibliography]{revtex4-2}
\usepackage[T1]{fontenc}

\usepackage{amsmath,amssymb}
\usepackage{colordvi}
\usepackage{graphicx}

\usepackage{xcolor}

\begin{document}

\title{Quantification of Quantum Correlations in Two-Beam Gaussian States Using Photon-Number Measurements}
\author{Artur Barasi\'nski$^{a,b}$}\email{artur.barasinski@uwr.edu.pl}
\author{Jan Pe\v{r}ina Jr.$^{a}$} \email{jan.perina.jr@upol.cz}
\author{{Anton\'\i n} \v{C}ernoch$^{a}$}
 \affiliation{$^{a}$Joint Laboratory of Optics of Palack\'{y}
University and Institute of Physics of CAS, Faculty of Science,
Palack\'{y} University, 17. listopadu 12, 771 46 Olomouc, Czech
Republic}
 \affiliation{$^{b}$Institute of Theoretical Physics,
Uniwersity of Wroclaw, Plac Maxa Borna 9, 50-204 Wroc\l aw,
Poland}

\begin{abstract}
Identification, and subsequent quantification of quantum
correlations, is critical for understanding, controlling, and
engineering quantum devices and processes. We derive and implement
a general method to quantify various forms of quantum correlations
using solely the experimental intensity moments up to the fourth
order. This is possible as these moments allow for an exact
determination of the global and marginal impurities of two-beam
Gaussian fields. This leads to the determination of steering,
tight lower and upper bounds for the negativity, and the
Kullback-Leibler divergence used as a quantifier of state
nonseparability. The principal squeezing variances are determined
as well using the intensity moments. The approach is demonstrated
on the experimental twin beams with increasing intensity and the
squeezed super-Gaussian beams composed of photon pairs. Our method
is readily applicable to multibeam Gaussian fields to characterize
their quantum correlations.
\end{abstract}

\date{\today}

\maketitle

Quantum theory allows for correlations between spatially separated
systems or degrees of freedom that are fundamentally different
from their classical counterparts. For composite systems, quantum
correlations (QCs) manifest themselves in many different
(inequivalent) forms
\cite{Weedbrookrmp84_2012,Quintinopra92_2015,Adesso_2016,De_Chiara_2018},
including the Bell nonlocality, quantum steering, and
entanglement. They can be exploited to achieve qualitatively
better performance in information processing tasks compared to
purely classical scenarios. Though the question of which kind of
QCs is a necessary resource for a given quantum information
protocol remains still open, particularly when multipart systems
are considered, the analysis of QCs among different subsystems is
extraordinarily important
\cite{Barasinskiprl122_2019,Barasinskisr8_15209_2018}. It belongs
to fundamental problems in quantum information science and quantum
many-body physics at present.
Applying very general quantum theories and models, the structure   
of QCs is elucidated and nontrivial limitations on the strength of
physically allowed QCs are revealed.

Over the years, several methods for the analysis of QCs have been
proposed based on, e.g., the violation of various inequalities
\cite{Chenprl88_2002,Shchukinprl95_2005,Cavaillesprl121_2018,BarasinskiQuantum_2021},
geometrical considerations \cite{Laskowskipra88_2013}, or even
interference among multiple copies of the investigated state
\cite{Horodeckipra68_2003,Fiurasek2004,Bartkiewiczpra95_2017}.
However, such methods usually require certain initial knowledge
about the analyzed state density matrix, which is never
experimentally acquired without technical difficulties.
Similarly, the homodyne tomography \cite{Lvovskyrmp81_2009} in
quantum optical experiments with continuous-variable states, which
provides the complete characterization of the detected field, has
to rely on a coherent local oscillator with the varying phase
\cite{Haderka09}. Contrary to this, we routinely measure the
photocount distributions in numerous experiments by
photon-number-resolving detectors \cite{Mandel1995}. So it is very
important to develop methods allowing to extract the maximum
information about QCs in the analyzed field using just these
photoncount distributions. While there are numerous
nonclassicality witnesses at our disposal \cite{PerinaJr2017a,
PerinaJr2020a, PerinaJr2022}, revealing the structure of QCs
represents a much more complicated task. Several schemes for
solving this problem have already been suggested using, however,
some form of homodyning of the analyzed field
\cite{Arkhipov2016a,Kuhn2017,Arkhipov2018b}. Interestingly, an
exact copy of the analyzed state can be used in a suitable
interferometric setup~\cite{Fiurasek2004} instead of a local
coherent oscillator of the homodyne scheme to reveal all four
invariants of a two-beam Gaussian field, which is equivalent to
the determination of all elements of its covariance matrix.


In this Letter, we address this problem for a wide group of
in-practice important Gaussian fields. We propose a scheme based
solely on the intensity moments of optical fields to estimate
their global and marginal purities. Although the photocount
measurements do not allow for the determination of all
coefficients characterizing such Gaussian fields (information
about the phases of complex coefficients is missing), the needed
information can partly be inferred from the values of higher-order
intensity moments \cite{Perina1991}. In our scheme, both the
global and marginal field purities are expressed in terms of
higher-order intensity moments. This opens the door for direct
determination of important QC quantifiers, involving the
R\'{e}nyi-2 entropy, Kullback-Leibler divergence, one- and two-way
Gaussian steering, and even tight lower and upper bounds for the
logarithmic negativity. Also, the squeezing of Gaussian states can
be determined. In practice, direct measurements of the intensity
moments of optical fields are standardly performed using various
types of photon-number-resolving detectors including intensified
CCD cameras \cite{Haderka2005a,MachulkaOE22_2014} or
superconducting bolometers \cite{Harder2016}, to name a few. This
makes our scheme qualitatively simpler compared to those based on
various forms of homodyning. Using two-beam states originated in
parametric down-conversion, we experimentally demonstrate the
suggested approach.


\textit{Purity estimation of Gaussian fields.---} We begin with
defining the normal characteristic function $C_{\cal N}(\beta_1,
\beta_2) $ \cite{Perina1991} for general single-mode two-beam
fields,
\begin{equation}  
 C_{\cal N}(\beta_1,\beta_2) = \langle \exp\Bigg[\sum_{j=1,2}
 \beta_j\hat{a}_j^\dagger\Bigg] \exp\Bigg[-\sum_{j=1,2}
 \beta_j^*\hat{a}_j\Bigg] \rangle,
\label{1}
\end{equation}
where $ \hat{a}_j $ ($ \hat{a}_j^\dagger $) stands for the
annihilation (creation) operator of beam $ j $, $ j=1,2 $ and $
\langle \cdots \rangle $ denotes quantum-mechanical averaging. For
quantum Gaussian fields the characteristic function $C_{\cal
N}(\beta_1, \beta_2)$ takes the form \cite{Perina1991}
\begin{eqnarray}   
\label{2}
 C_{\mathcal{N}}(\beta_1,\beta_2,\beta_1^*,\beta_2^*) &=& \exp \Bigg[-\sum_{j=1,2} [ B_j |\beta_j|^2
    \\
   + ( C_j \beta^{* 2}_j/2 + {\rm c.c.} )] &+& (D_{12} \beta^*_1 \beta^*_2 +
   \bar{D}_{12} \beta_1 \beta^*_2 + {\rm c.c.} \big) \big] \nonumber
\end{eqnarray}
with real ($B_j $) and complex ($ C_j $, $ D_{jk} $, $
\bar{D}_{jk} $) parameters characterizing the Gaussian state;
${\rm c.c.} $ replaces the complex-conjugated terms.

The measured intensity moments $ \langle W_1^k W_2^l\rangle$,
given as the normally ordered photon-number moments and obtained
via the Stirling numbers from the measured photon-number
moments~\cite{Perina1991}, contain partial information about the
state parameters:
\begin{eqnarray} 
&\langle W_1^k W_2^l\rangle = (-1)^{k+l}
 \frac{\partial^{2(k+l)}C_{\mathcal{N}}(\beta_1,\beta_2,\beta_1^*,\beta_2^*)}{\partial^k\beta_1
 \partial^k(\beta_1^*)\partial^l\beta_2
 \partial^l(\beta_2^*)}\bigg|_{\beta_1 =... = \beta_2^* = 0}. & \nonumber \\
 & & \label{3}
\end{eqnarray}

Considering the intensity moments up to the second order, we
reveal the following relations for the looked-for parameters:
\begin{eqnarray}  
 B_j &=& \langle W_j \rangle, \nonumber \\
 |C_j|^2 &=& \langle W_j^2\rangle - 2 \langle W_j\rangle^2, \hspace{5mm} j=1,2, \nonumber \\
 |D_{12}|^2+|\bar{D}_{12}|^2 &=& \langle W_1W_2\rangle - \langle W_1\rangle \langle W_2\rangle.
\label{4}
\end{eqnarray}
The third-order intensity moments give us additional information
about the looked-for parameters:
\begin{eqnarray}  
 -4 \Re \{C_1 \bar{D}_{12} D^*_{12}\} &=& -4 \langle W_1 \rangle \Big(\langle W_1W_2 \rangle-\langle W_1\rangle \langle W_2\rangle\Big)\nonumber\\
 &+&\langle W_1^2W_2\rangle - \langle W_1^2\rangle \langle W_2\rangle, \nonumber \\
 -4 \Re \{C^*_2 \bar{D}_{12} D_{12}\} &=& -4 \langle W_2 \rangle \Big(\langle W_1W_2 \rangle-\langle W_1\rangle \langle W_2\rangle\Big)\nonumber \\
 &+& \langle W_1W^2_2\rangle - \langle W_1\rangle \langle W_2^2\rangle.
\label{5}
\end{eqnarray}
We note that, alternatively, the expressions in Eq.~(\ref{5}) can
be obtained from the third-order moments $ \langle W_j^3\rangle $,
$j=1,2 $ or even fourth-order moments $ \langle W_1^3W_2\rangle $
and $ \langle W_1W^3_2\rangle $. Finally, also the fourth-order
intensity moment $ \langle W_1^2W_2^2\rangle $ reveals a useful
relation among the looked-for parameters,
\begin{eqnarray}  
 & 4 (2 |D_{12}|^2 |\bar{D}_{12}|^2 + \Re \{C_1 C^*_2 \bar{D}^2_{12}\}
  +\Re \{C_1 C_2 (D^*_{12})^2\} ) = & \nonumber \\
 & \langle W_1^2W_2^2\rangle
  - 4 \langle W_1W_2\rangle^2 - \langle W_1\rangle \langle W_2\rangle
  \langle W_1W_2\rangle  + \langle W_1\rangle^2 \langle W_2\rangle^2& \nonumber\\
 &  -2 [\langle W_1\rangle^2 (\langle W_2^2\rangle - 2 \langle W_2\rangle^2) +\langle W_2\rangle^2 (\langle W_1^2\rangle - 2 \langle W_1\rangle^2)]& \nonumber \\
 &  - (\langle W_2^2\rangle - 2 \langle W_2\rangle^2)~(\langle W_1^2\rangle - 2 \langle W_1\rangle^2)& \nonumber\\
 & + 16 \langle W_2\rangle \Re \{C_1 \bar{D}_{12} D^*_{12}\} +16 \langle W_1\rangle \Re \{C^{\ast}_2 \bar{D}_{12} D_{12}\}.&
\label{6}
\end{eqnarray}
Surprisingly, Eqs.~(\ref{4})--(\ref{6}) are sufficient to obtain
the global and marginal purities of the two-beam Gaussian fields.
In order to show that, we need to know the corresponding
covariance matrix $\boldsymbol{\sigma} \equiv \langle
\hat{\boldsymbol \xi} \hat{\boldsymbol \xi}^{T} \rangle-\langle
\hat{\boldsymbol \xi} \rangle \langle \hat{\boldsymbol \xi}^{T}
\rangle$ containing the second-order moments of the position $
\hat{x}_j = (\hat{a}_j + \hat{a}_j^\dagger)/2 $ and momentum $
\hat{p}_j =(\hat{a}_j - \hat{a}_j^\dagger)/(2i) $ operators
embedded in vector $ \hat{\boldsymbol \xi}^{T} =
(\hat{x}_1,\hat{p}_1,\hat{x}_2,\hat{p}_2) $. Using the parameters
of the normal characteristic function $ C_{\cal N} $ in
Eq.~(\ref{2}), the covariance matrix $ \boldsymbol \sigma $ is
expressed as
\begin{eqnarray}   
 {\boldsymbol \sigma} &=& \left[ \begin{array}{cc} {\boldsymbol \sigma}_1 & {\boldsymbol \gamma} \\
  {\boldsymbol \gamma}^T & {\boldsymbol \sigma}_2 \end{array} \right],
\label{7}
\end{eqnarray}
where
\begin{eqnarray}   
 {\boldsymbol \sigma}_j &=& \left[ \begin{array}{cc}
  1+2 B_j+2 \Re \{C_j\} & 2 \Im \{C_j\} \\
   2 \Im \{C_j\} & 1+2 B_j-2 \Re \{C_j\} \end{array} \right],  \nonumber \\
 {\boldsymbol \gamma} &=& \left[ \begin{array}{cc}
  2 \Re \{D_{12}-\bar{D}_{12}\} & 2 \Im \{D_{12}-\bar{D}_{12}\} \\
  2 \Im \{D_{12}+\bar{D}_{12}\} & -2 \Re \{D_{12}+\bar{D}_{12}\} \end{array} \right].
\label{8}
\end{eqnarray}

Now, one can easily verify that the determinants of the global $
\boldsymbol \sigma $ and local $ {\boldsymbol \sigma}_j $, $ j=1,2
$, covariance matrices are given in terms of the intensity moments
as
\begin{eqnarray}   
 \det{\boldsymbol \sigma} &=& 1 + 4 (\langle W_1 \rangle + \langle W_2\rangle ) + 12 (\langle W_1 \rangle + \langle W_2\rangle )^2
  \nonumber \\
 & & - 4 \langle W_1^2 \rangle (1 + 6 \langle W_2\rangle + 24 \langle W_2\rangle^2 ) - 4 \langle W_2^2\rangle  \nonumber\\
 & & \times (1 + 6 \langle W_1 \rangle + 24 \langle W_1 \rangle^2 ) + 8 \langle W_1^2W_2\rangle  \nonumber \\
 & & \times (1 + 6 \langle W_2\rangle) + 8 \langle W_1 W_2^2\rangle (1 + 6 \langle W_1 \rangle ) \nonumber\\
 & &  \hspace{-5mm} - 8 \langle W_1 W_2\rangle (1 + 6 \langle W_1 \rangle + 6 \langle W_2\rangle + 48 \langle W_1 \rangle \langle W_2\rangle ) \nonumber \\
 & & + 96 \langle W_1 \rangle \langle W_2\rangle (\langle W_1 \rangle + \langle W_2\rangle + 5 \langle W_1 \rangle \langle W_2\rangle ) \nonumber\\
 & & \hspace{-5mm}  + 24 \langle W_1^2 \rangle \langle W_2^2\rangle
  - 8 \langle W_1^2 W_2^2\rangle + 48 \langle W_1W_2\rangle^2,
\label{9}  \\
 \det{\boldsymbol \sigma}_j &=& 1+4 \langle W_j\rangle + 12\langle W_j\rangle^2 - 4 \langle W_j^2\rangle, j=1,2.
\label{10}
\end{eqnarray}
Knowing these determinants, the corresponding purities $ \mu
=1/(\det{\boldsymbol \sigma})^{1/2}$ and $ \mu_j=
1/(\det{\boldsymbol \sigma}_j)^{1/2}$ are established
\cite{Weedbrookrmp84_2012}. Contrary to this, seralian $ \Delta $,
the last of four global invariants of two-beam Gaussian states,
requires $ {\boldsymbol \gamma} $ to be determined, $ \Delta =
\det{\boldsymbol \sigma}_1 + \det{\boldsymbol \sigma}_2 + 2
\det{\boldsymbol \gamma} $ \cite{McHugh2006,Adesso_2016}.

This central result allows us to determine various quantities that
characterize the structure of two-beam QCs
\cite{Hill1997,Cavalcanti2009,Adessoprl109_2012,Kogias2015}. Using
the purities $\mu$ and $\mu_j$, $ j=1,2 $, we immediately obtain
the R\'{e}nyi-2 entropies along the formula $ S_R = - \ln (\mu)$
\cite{Adessoprl109_2012}. We note that $ S_R $ represents the
continuous analog of the Shannon entropy. The R\'{e}nyi-2 entropy
$ S_R $ can then be used to quantify the total quadrature
correlations via the Kullback-Leibler divergence (distance) $ H $
between the analyzed two-beam state $ \hat{\rho} $ and its
factorized counterpart $ {\rm Tr}_2\{\hat{\rho}\} {\rm
Tr}_1\{\hat{\rho}\} $ \cite{Adessoprl109_2012}:
\begin{equation}  
 H = S_{R,1} + S_{R,2} - S_R = \ln \left(\frac{\mu}{\mu_1 \mu_2}\right).
\label{11}
\end{equation}
Also the degree of (one-way) Gaussian steering of beam $ 2 $ by
beam $ 1 $ \cite{Cavalcanti2009} is expressed in terms of purities
\cite{Kogias2015}:
\begin{equation}  
 {G}_{1 \rightarrow 2} = {\rm max} \{0, \ln(\mu/\mu_1)\}.
\label{12}
\end{equation}
We note that two-way steering occurs provided that both $ {G}_{1     
\rightarrow 2}$ and $ {G}_{2 \rightarrow 1} $ are nonzero.

Using purities, even the logarithmic negativity $ E_N $
\cite{Hill1997}, giving the degree of entanglement revealed, e.g.,
by the Simon criterion \cite{Simon2000}, is established through
its tight lower and upper bounds derived by Adesso \emph{et al.}
\cite{Adesso2004}:
\begin{eqnarray}  
 E_{\max}(\rho) & = &-\frac{1}{2} \ln \bigg[-\frac{1}{\mu}
  +\left(\frac{\mu_1+\mu_2}{2 \mu_1^2 \mu_2^2}\right) \nonumber \\
  &&\bigg(\mu_1 + \mu_2-\sqrt{(\mu_1+\mu_2)^2-\frac{4 \mu_1^2 \mu_2^2}{\mu}}\bigg)\bigg], \nonumber\\
 E_{\min}(\rho)&=&-\frac{1}{2} \ln \bigg[\frac{1}{\mu_1^2}+\frac{1}{\mu_2^2}-\frac{1}{2
  \mu^2} - \frac{1}{2} \nonumber \\
  &-&\sqrt{\bigg(\frac{1}{\mu_1^2}+\frac{1}{\mu_2^2}-\frac{1}{2
  \mu^2}-\frac{1}{2}\bigg)^2-\frac{1}{\mu^2}}\bigg].
\label{13}
\end{eqnarray}

\begin{figure}  
  \centerline{\includegraphics[width=0.47\hsize]{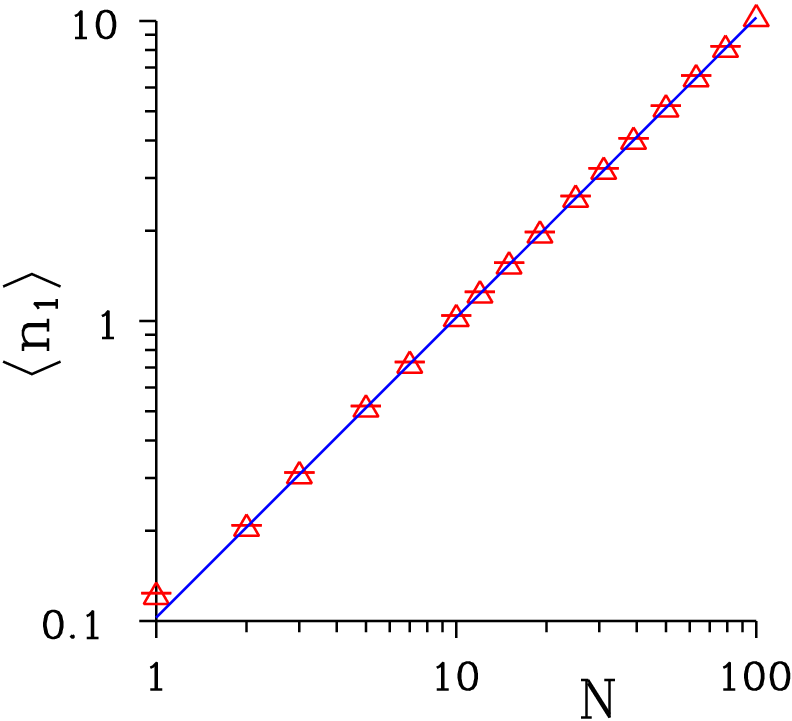}
    \hspace{2mm}
     \includegraphics[width=0.47\hsize]{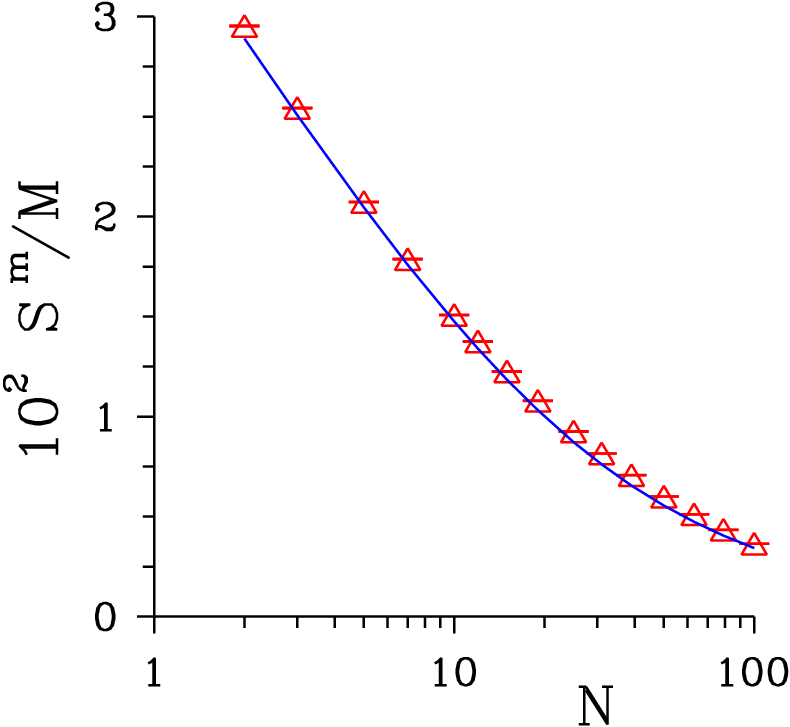}}
  \centerline{(a)\hspace{0.45\hsize} (b)}
  \centerline{\includegraphics[width=0.47\hsize]{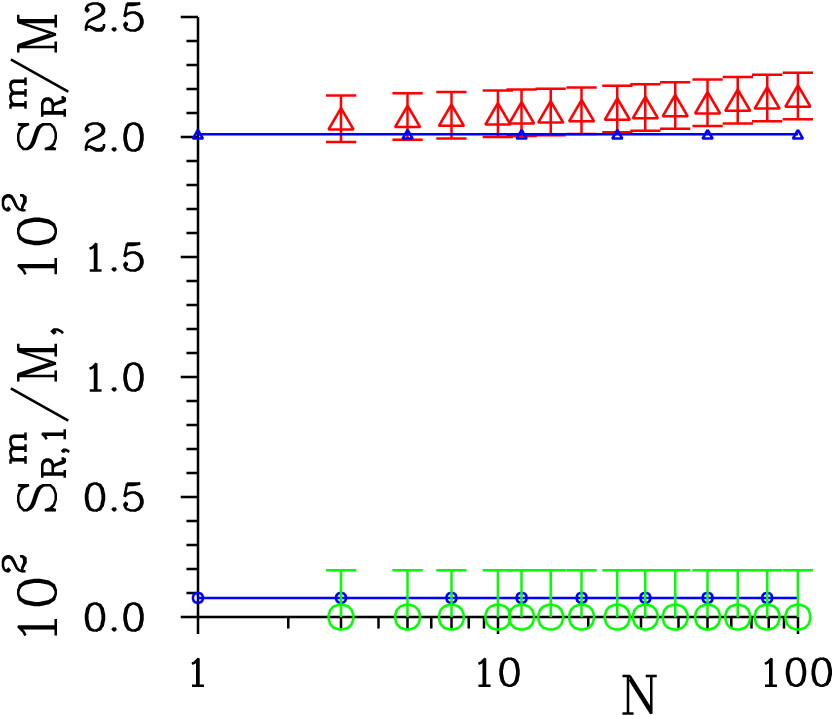}
    \hspace{2mm}
     \includegraphics[width=0.47\hsize]{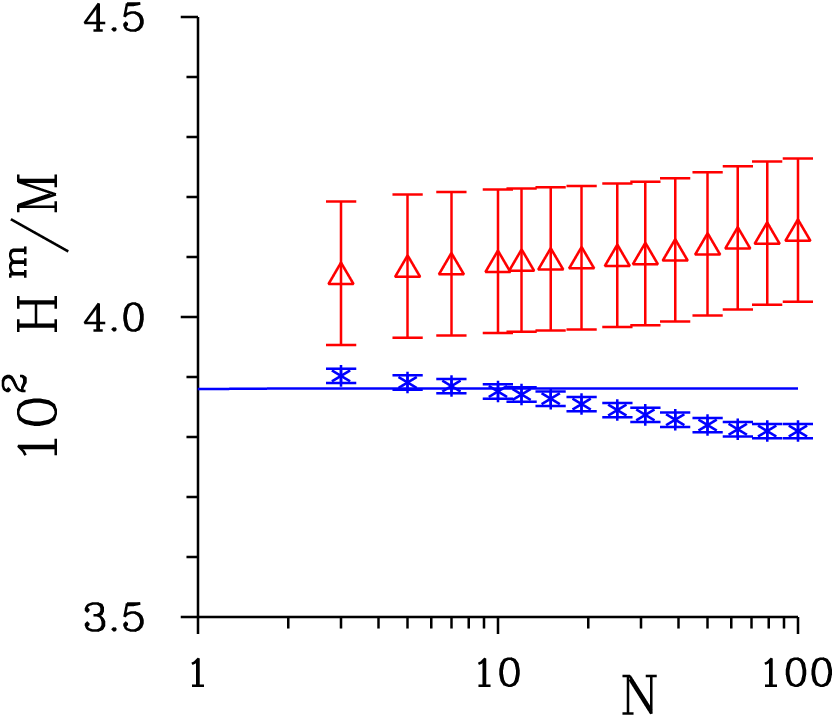}}
  \centerline{(c)\hspace{0.45\hsize} (d)}
  \centerline{\includegraphics[width=0.47\hsize]{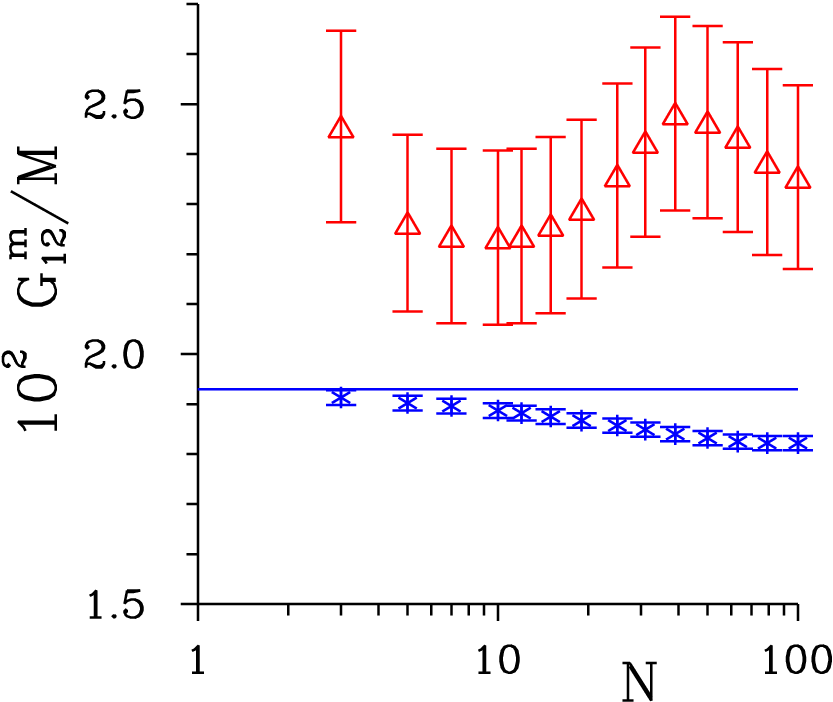}
    \hspace{2mm}
     \includegraphics[width=0.47\hsize]{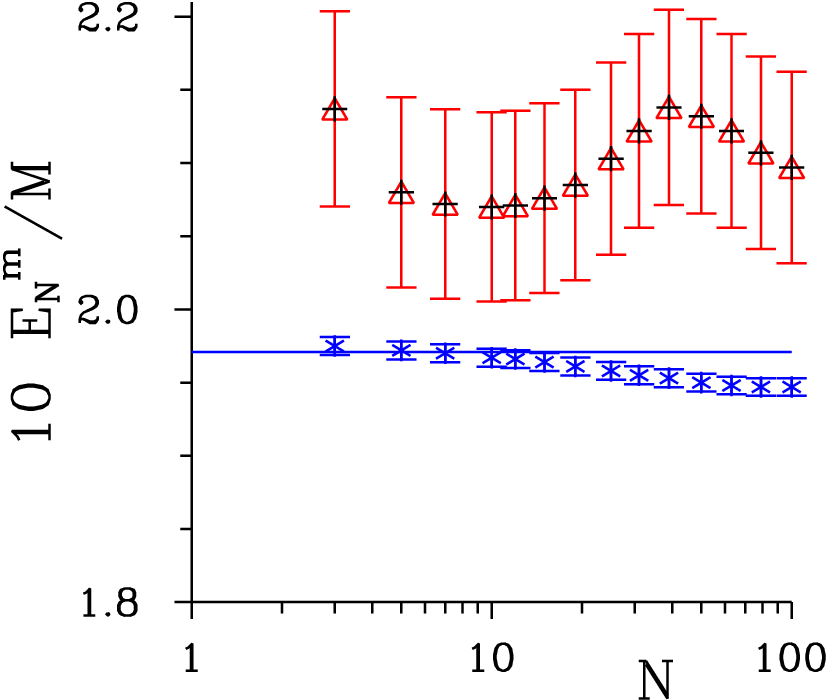}}
  \centerline{(e)\hspace{0.45\hsize} (f)}
 \caption{(a) Mean photon number $ \langle n_1\rangle $ of beam 1 (red $ \triangle $),   
   (b) entropy $ S^{\rm m}/M $ per mode (red $ \triangle $), (c)
   R\' enyi-2 entropies per mode $ S_{R,1}^{\rm m}/M $ and $ S_{R}^{\rm m}/M $
   for beam 1 (red $ \triangle $) and beams 12 (green $ \circ
   $), respectively, (d) Kullback-Leibler divergence per mode $ H^{\rm m}/M $
   and its values for noisy TWBs (blue $ \ast $), (e) steering parameter
   per mode $ G_{12}^{\rm m}/M $ (red $ \triangle $) and its values for noisy TWBs (blue $ \ast $),
   and (f) lower (red $ \triangle $) and upper (black +)
   bounds for negativity per mode $ E_N^{\rm m}/M $ and its values for
   noisy TWBs (blue $ \ast $) as they depend on the number $ N $ of grouped detection windows;
   $ M = 10 N $ (for details, see \cite{SM,PerinaJr2021b}).
   In (a) and (b) error bars are smaller than the plotted symbols.
   In (f), red $ \triangle $ and black + nearly coincide.
   The solid blue curves originate in a model of $ M $ identical independent single-mode two-beam Gaussian
   fields with suitable parameters.}
\label{fig1}
\end{figure}

\textit{Application to experimental data.---} We tested the
derived formulas on a set of the experimental spatiospectrally
multimode twin beams (TWBs) \cite{PerinaJr2019} with increasing
intensity [for mean photon number $ \langle n_1\rangle $ of beam
1, see Fig.~\ref{fig1}(a)] that were obtained by adding the
photocounts registered by two single-photon counting modules
positioned in the signal (1) and idler (2) beams in subsequent
detection windows (for details, see the Supplemental Material
\cite{SM} and \cite{PerinaJr2021b}). The TWBs at 710 nm originated
in type-I parametric down-conversion in a LiIO$ {}_3 $ nonlinear
crystal pumped by the third harmonic of an Nd-YAG laser at 355 nm.
We arrived this way at the compound multimode TWBs with mean
photon numbers extending over 2 orders in magnitude (from 0.1 to
10 mean photons per beam). The experimental photocount histograms
were reconstructed by the maximum-likelihood approach to obtain
the joint photon-number distribution $ p(n_1,n_2) $ and its photon
number moments $ \langle n_1^k n_2^l \rangle_{\rm m} =
\sum_{n_1,n_2=0}^{\infty} n_1^k n_2^l p(n_1,n_2) $. Also entropy $
S $ of the fields was determined along the formula $ S = -
\sum_{n_1,n_2=0}^{\infty} p(n_1,n_2) \ln[p(n_1,n_2)] $ and plotted
in Fig.~\ref{fig1}(b). The intensity moments, that are the
normally ordered photon-number moments, were then derived as
linear combinations of photon-number moments using the Stirling
numbers of the first kind \cite{PerinaJr2017a}.

We note that, when correcting the experimental data for nonunit
detection efficiencies $ \eta_1 $ and $ \eta_2 $, the use of any
reconstruction method is not required. We may simply determine the
intensity moments $ \langle W_{1}^k W_{2}^l\rangle_E $ directly
from the detected photocount histogram and derive the needed
intensity moments as $ \langle W_{1}^k W_{2}^l\rangle = \langle
W_{1}^k W_{2}^l\rangle_E / (\eta_1^k\eta_2^l) $. On the other
hand, reconstruction methods allow us to correct also for other
detector parameters like dark-count rates, cross talk, etc. In
real experiments, sufficiently large detection efficiencies are
needed to arrive at acceptably low errors in the determination of
the reconstructed intensity moments.

These experimental beams are multimode as they are generated in
parametric down-conversion in the running-wave configuration and
detected in multiple detection windows. They also suffer from
imperfections that occur both during the generation process and
transmission to the detectors. As a consequence, they decline from
the theoretically expected form of a multimode noisy TWB whose
coefficients obey $ C_1 = C_2 = \bar{D}_{12} = 0 $ in
Eq.~(\ref{2}). These declinations can be quantified using the
derived formulas for the tested quantum information quantities.
However, these quantities are derived for two single-mode Gaussian
beams and so their application is conditioned by the reduction of
the experimental multimode photon-number moments $ \langle n_1^k
n_2^l \rangle_{\rm m} $ to one typical mode in each beam.
Estimating the number $ M $ of effective modes in each
beam~\cite{PerinaJr2022,Michalek2020} we may follow the procedure
outlined in the Supplemental Material \cite{SM}.

We compare the values of the obtained quantities per mode with
those characterizing a single-mode Gaussian noisy TWB
\cite{Arkhipov2015}, that is fully characterized by three first-
and second-order photon-number moments ($ B_j = \langle w_j
\rangle = \langle n_j \rangle$, $ |D_{12}|^2 = \langle w_1 w_2
\rangle - \langle w_1 \rangle \langle w_2 \rangle = \langle n_1
n_2 \rangle - \langle n_1 \rangle \langle n_2 \rangle $). For the
formulas, see the Supplemental Material \cite{SM}.

According to the experimental results reduced to one mode and
plotted in Figs.~\ref{fig1}(c)---(f), the R\' enyi-2 entropies $
S_R^{\rm m}/M $, the Kullback-Leibler divergence $ H^{\rm m}/M $,
the negativity $ E_N^{\rm m}/M $ as well as the steering parameter
$ G_{1\rightarrow 2}^{\rm m}/M $ do not considerably change with
the increasing field intensity, i.e., the increasing number $ N $
of grouped detection windows ($ M = 10 N $).

As the values of R\' enyi-2 entropy $ S_{R}^{\rm m} $ of the
two-beam fields are smaller than the entropies $ S_{R,1}^{\rm m} $
and $ S_{R,2}^{\rm m} $ of the constituting signal and idler beams
[see Fig.~\ref{fig1}(c)], the purities of the two-beam fields are
greater than those of the constituting beams. This implies,
according to the general classification of two-beam Gaussian
states (see Table~I in \cite{Adessopra70_2004}), that the analyzed
two-beam fields are entangled.

The experimental values of 
$H_{R}^{\rm m}/M $, 
$G_{1\rightarrow 2}^{\rm m}/M $ and 
$E_{N}^{\rm m}/M $ reduced per one mode and determined by the
derived formulas (\ref{11})---(\ref{13}) are systematically
greater (by approximately 10\% -- 20\%) than those characterizing
the Gaussian noisy TWBs (determined by the formulas in the
Supplemental Material \cite{SM}). This means that the states of
the detected two-beam fields are more general than those of the
Gaussian noisy TWBs with the vanishing coefficients $ C_1 $, $ C_2
$, and $ \bar{D}_{12} $. The consideration of the experimental
third- and fourth-order intensity moments reveals that also the
complex parameters ($ C_1 $, $ C_2 $, and $ \bar{D}_{12} $) of the
detected two-beam Gaussian fields are nonzero, which leads,
according to our results, to stronger QCs described by the above
quantities. We note here, that our results do not allow us to
judge the declination (non-Gaussianity) of the analyzed state from
the general form of Gaussian states as described by the
characteristic function in Eq.~(\ref{2}).

The Kullback-Leibler divergence $ H_{R}^{\rm m} $, the steering
parameter $ G_{1\rightarrow 2}^{\rm m} $ and the negativity $
E_{N}^{\rm m} $ of the two-beam fields increase practically
linearly with the increasing TWB intensity. This contrasts with
the behavior of the entropy $ S^{\rm m} $ of the two-beam fields
whose increase is smaller: The entropy $ S^{\rm m}/M $ per mode
plotted in Fig.~\ref{fig1}(b) decreases with the increasing TWB
intensity. This means that the capacity of available QCs increases
linearly with the dimensionality (number of modes $ M $)
of the analyzed fields. The capacity of QCs thus grows faster than       
disorder in the analyzed fields quantified by the entropy $ S^{\rm
m} $.

\begin{figure}  
  \centerline{\includegraphics[width=0.47\hsize]{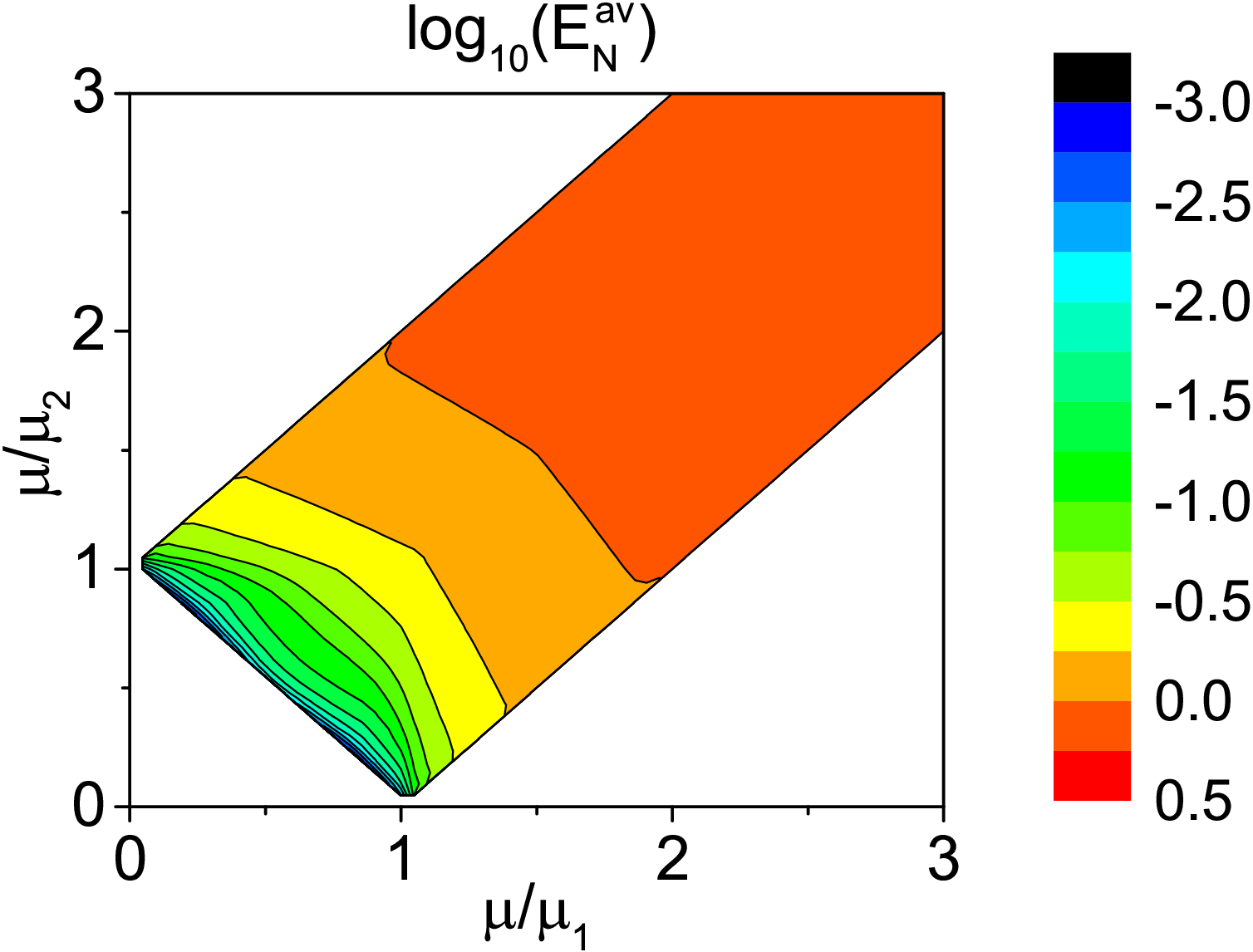}
    \hspace{2mm}
     \includegraphics[width=0.47\hsize]{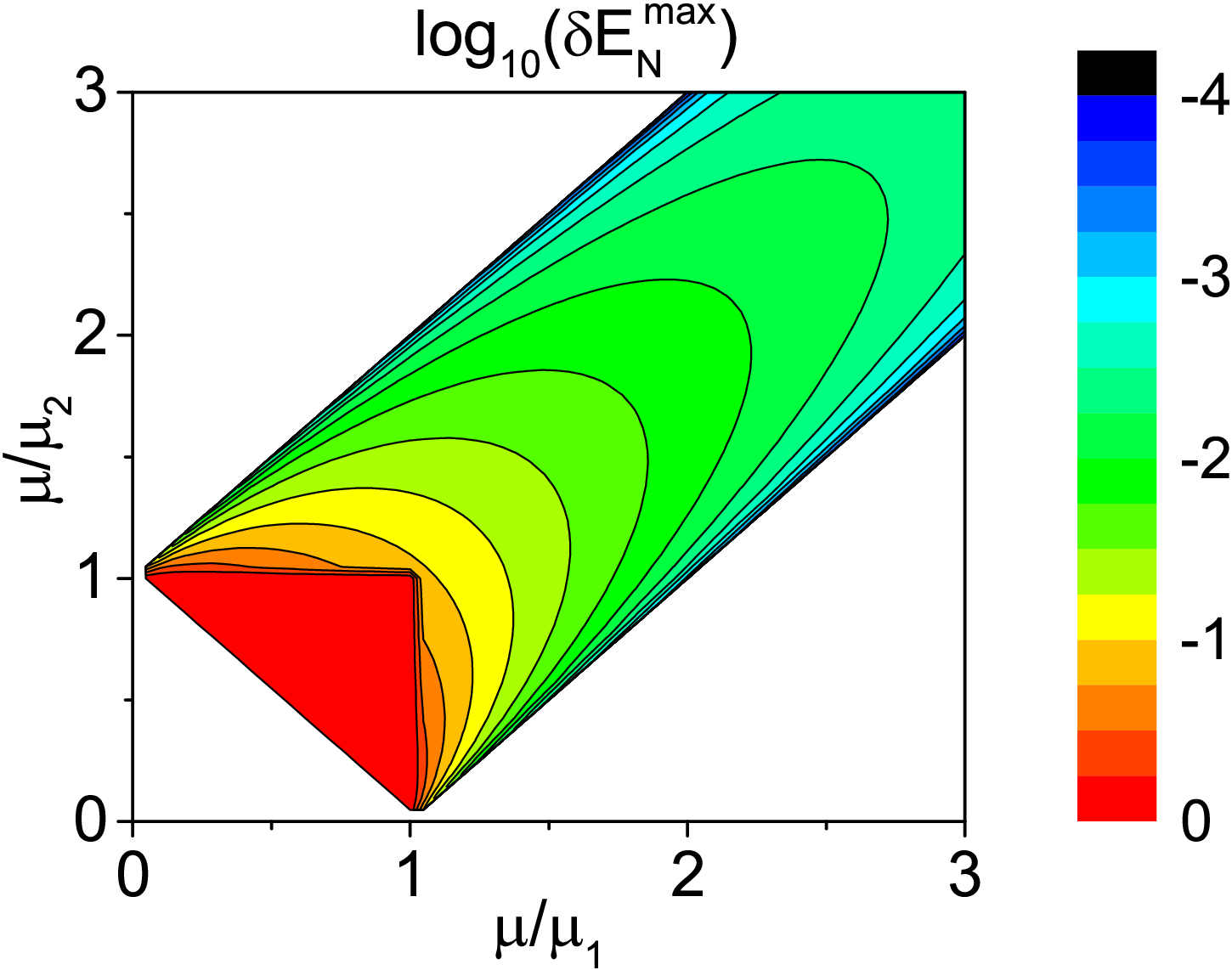}}
  \centerline{(a)\hspace{0.45\hsize} (b)}

 \caption{(a) Average negativity $ E_N^{\rm av} $ and (b) maximum $ \delta
   E_N^{\rm max} $ of relative error as they depend on ratios $ \mu/\mu_1 $
   and $ \mu/\mu_2 $. In the white areas no entangled states
   exist. For $ \mu_1=\mu_2 $, see also in~\cite{Kogias2015}.}
\label{fig2}
\end{figure}

We note that nonzero negativity $ E_{N}^{\rm m} $ and the
Kullback-Leibler divergence $ H_{R}^{\rm m} $ are obtained also
directly for the experimental photocount histograms, i.e., without
reconstructing the experimental data. This contrasts with the
steering parameters $ G_{1\rightarrow 2}^{\rm m} $ and $
G_{2\rightarrow 1}^{\rm m} $ being zero in this case.

The negativity $ E_N $ is the most commonly used parameter to
quantify QCs. However, we have only the lower and upper bounds in
Eq.~(\ref{13}) at disposal for single-mode two-beam Gaussian
fields. Nevertheless, the experimental data plotted in
Fig.~\ref{fig1}(f) show that these bounds are very close to each
other: The uncertainty in determining $ E_N $ is practically given
only by the experimental errors. This observation is valid in
general. Indeed, from the point of view of the entanglement,
two-beam Gaussian states are divided into groups of states with
identical amount of $ E_N $. These groups are parameterized by
four parameters: purities $ \mu $, $ \mu_1 $, $ \mu_2 $, and
seralian $ \Delta $. The minimal and maximal values of the
negativity $ E_N $ given in Eq.~(\ref{13}) in fact represent the
extremal values with respect to seralian $ \Delta $ for fixed
values of the purities. The general behavior of these extremal
values can conveniently be quantified taking into account that the
negativity $ E_N $ increases (decreases) with the increasing
global purity $ \mu $ (marginal purities $ \mu_1 $ and $ \mu_2 $).
This suggests the ratios $ \mu/\mu_j $, $ j=1,2 $, as suitable
parameters for quantifying the uncertainty in the determination of
$ E_N $: The greater the ratios are, the greater the negativity $
E_N $ is. This is documented in the graph of Fig.~\ref{fig2}(a)
where the negativity $ E_N^{\rm av} $ averaged over the states
with fixed ratios $ \mu/\mu_1 $ and $ \mu/\mu_2 $ is plotted. The
maximum of the relative error $ \delta E_N^{\rm max} $
\cite{Adesso2004},
\begin{equation}  
 \delta E_N = \frac{ E_N^{\rm max} - E_N^{\rm min} }{
  E_N^{\rm max} + E_N^{\rm min} } ,
\label{14}
\end{equation}
taken over the states with the fixed ratios $ \mu/\mu_1 $ and $
\mu/\mu_2 $ is then shown in Fig.~\ref{fig2}(b). According to
Fig.~\ref{fig2}(b), the relative error $ \delta E_N  $ is smaller
than 10\% (1\%) when the ratios $ \mu/\mu_j $ are greater than 1.5
(2.5), i.e., when the states exhibit considerable entanglement.
This makes the use of the bounds for negativity $ E_N $ very
efficient.

\textit{Single beam properties.---} The approach presented above
for two-beam Gaussian fields is applicable also to single-beam
Gaussian fields. The intensity moments allow us to determine the
principal squeezing variance \cite{Luks1988,Dodonov2002} in this
case. Merging the intensities of the signal and idler beams of the
above discussed two-beam fields, we arrive at single-beam
super-Gaussian fields with phase fluctuations reduced below the
shot-noise limit, as discussed in detail in the Supplemental
Material \cite{SM}.

\textit{Further extension and application.---} Our results also
allow for the analysis of QCs of general $n$-beam Gaussian states.
This is so as the appropriate covariance matrix is composed of
blocks of $ 2\times 2$ matrices similar to those written in
Eq.~(\ref{7}) \cite{Adessoprl93_2004,Adessopra73_2006}. This
allows us to analyze its properties by considering all possible
two-beam subsystems of the whole $n$-beam Gaussian field. The
formulas in Eqs.~(\ref{9}) and (\ref{10}) hold for such subsystems
and allow us to quantify QCs in these two-beam reductions. Relying
on various monogamy relations, we may establish the lower bound
for the genuine multipartite QCs in the whole $n$-beam field
\cite{Adessoprl109_2012,Lamiprl117_2016,Xiangpra95_2017}.

In {\it conclusion}, we have shown how various forms of quantum correlations 
of two-beam Gaussian fields (with spatiospectral multimode
structure), that naturally depend on the fields phase properties,
can be quantified solely from the measured intensity moments up to
the fourth order. The determination of the global and marginal
purities of the involved beams in terms of the intensity moments
represents the key step. The principal squeezing variances can
solely be derived from the intensity moments as well. We have
demonstrated usefulness and practicality of this approach by
considering suitable experimental fields. Our method is readily
applicable to multipartite systems for the detection and
characterization of their quantum correlations. As the Gaussian
states are exploited in numerous metrology applications and
quantum-information protocols with continuous variables, we
foresee numerous applications of the suggested and demonstrated
method in the near future.

\acknowledgements{A.B. and A.\v{C}. acknowledge financial support
by the Czech Science Foundation under the project No.~20-17765S.
The authors thank the project CZ.02.1.01/0.0/0.0/16\_019/0000754
of the Ministry of Education, Youth and Sports of the Czech
Republic.}


%

\end{document}